# Acousto-optic laser optical feedback imaging

O. Jacquin[1,*], W. Glastre[1], E. Lacot[1], O.Hugon[1], H. Guillet de Chatellus[1] and F.Ramaz[2]

[1]Centre National de la Recherche Scientifique / Université de Grenoble 1, Laboratoire Interdisciplinaire de Physique, UMR 5588,Grenoble, F- 38041

[2] Institut Langevin "ondes et images, CNRS : UMR7587 – ESPCI ParisTech – Université Paris VI - Pierre et Marie Curie – Université Paris VII - Paris Diderot

*Corresponding author: Olivier.jacquin@ujf-grenoble.fr

We present a photon noise and diffraction limited imaging method combining the imaging laser and ultrasonic waves. The laser optical feedback imaging (LOFI) technique is an ultrasensitive imaging method for imaging objects through or embedded within a scattering medium. However, LOFI performances are dramatically limited by parasitic optical feedback occurring in the experimental setup. In this work, we have tagged the ballistic photons by an acousto-optic effect in order to filter the parasitic feedback effect and to reach the theoretical and ultimate sensitivity of the LOFI technique. We present the principle and the experimental setup of the acousto-optic laser optical feedback imaging (AO-LOFI) technique, and we demonstrate the suppression of the parasitic feedback.

OCIS Codes: OCIS Codes:, (110.4280) , (110.3175) , (280.3420)

## Context

Imaging objects through or embedded within a scattering media is a challenging problem linked to many medical applications such as cancer detection. The main challenge is to realize images through a turbid medium with both a high resolution and good signal to noise ratio (SNR). The information needed to obtain diffraction-limited images is carried by ballistic photons. However compared to scattered photons their number rapidly decreases with the depth, which dramatically reduces the SNR. Consequently, imaging through scattering media with diffraction-limited resolution requires both the detection of ballistic photons and the rejection of the scattered light. The filtering of the scattered light is generally achieved by limiting the depth of focus of the imaging setup. Efficient methods for accomplishing this goal include confocal microscopy [1], time resolved techniques [2] and optical coherent microscopy [3]. However, the thickness of the scattering medium explored with these techniques is limited to about twenty mean free paths and these methods require a high optical power in the medium to compensate for the losses in ballistic photons, which is often not compatible with medical application. To image deeper with a "good" SNR, there is a great interest in using the scattered photons since they decrease much slower than the ballistic ones. In this case, a millimetric resolution can be obtained by acoustic tagging of the scattered photon with a focalized ultrasonic wave, as in Acousto-Optical Coherence Tomography (AOCT) [4].

## LOFI technique and imaging through turbid media

We propose to use the LOFI technique while focalizing an ultrasonic (US) wave in a scattering media in order to image embedded object with a diffraction-limited resolution. LOFI is an ultrasensitive laser autodyne interferometer and also a confocal imaging technique combining the great accuracy of optical interferometry with the very high sensitivity of class B lasers to optical feedback [5].In this autodyne method, the optical beating between a reference wave and the signal wave (the light back-reflected by the target) takes place inside the laser source cavity, and can be amplified by the laser gain. Thanks to a resonant amplification, an optical feedbacks as low as -130dB is then detectable in a 1Khz detection bandwidth, with a laser output power of a few mW [6]. The confocal nature of the method is set by the coupling between the intracavity light mode and the light back-reflected by the target. Indeed, the fundamental transverse mode ($TEM_{00}$mode) of the laser plays the role of a spatial filter, like the pinhole of a confocal microscope. In the LOFI technique, the laser plays the role of both the emitter and of the detector, and the SNR of the method can easily be shot noise limited [7]. In conclusion, the LOFI method is *a priori* a powerful imaging method to realize simultaneously reflectivity and phase (i.e. profilometry) images under non-cooperative conditions such as imaging through a scattering media with a low optical power.

However, parasitic optical feedbacks present in the experimental setup dramatically limit the lowest detectable target reflectivity [8]. Indeed, it may be problematic to detect a target signal of lower magnitude than the noise reinjected by the parasitic feedback. We have proposed in a previous paper a solution to isolate the parasitic feedback effect based on a two beams setup, which unfortunately limits the numerical aperture of the focusing lens and significantly decreases the resolution [8]. To avoid the parasitic optical feedback effect while preserving the lateral optical resolution, we propose here to locally tag the photons of interest (i.e. back-reflected by the target) with a focused US wave and to detect selectively these photons by heterodyne filtering.

## LOFI setup with acoustic tagging

A scheme of our experimental setup is shown in Fig 1. First, we describe a conventional LOFI setupwithout local acoustic tagging (i.e. without the acoustic transducer of fig.1) which is thesame as one describedin [7]. The laser is a cw $Nd^{3+}$:YAG microchip with an output power $P_{out}$=40mW (i.e. $p_{out}$=2.14x$10^{17}$photon/s), at the wavelength λ=1064nm. The laser beam is frequency shifted through two acousto-optic modulators (AOM), the frequency shift is tunable and is noted $F_A/2$. Then the laser beam is sent onto the studied target using a two-axis galvanometric mirror scanner and a focusing lens. The back-reflectedlight by the target is reinjected in the laser by the same path. After a round-trip of the light, the frequency shift introduced by the AOMs is $F_A$. The photons re-injected inside the laser create an optical beating and lead to a modulation of the output power at the frequency $F_A$. There is a resonant amplification of this modulation if $F_A$ is close to the laser relaxation frequency $F_R$ and in this case, the method is *a priori* shot noise limited. A beam splitter sends a fraction of the output power on the photodiode connected to a lock-in amplifier which gives the magnitude and the phase of the output power modulation. The LOFI images are built point by point by moving the laser beam on the target. Then, it is possible to realize simultaneously reflectivity and phase images of the target in non-cooperative conditions such as through turbid media.

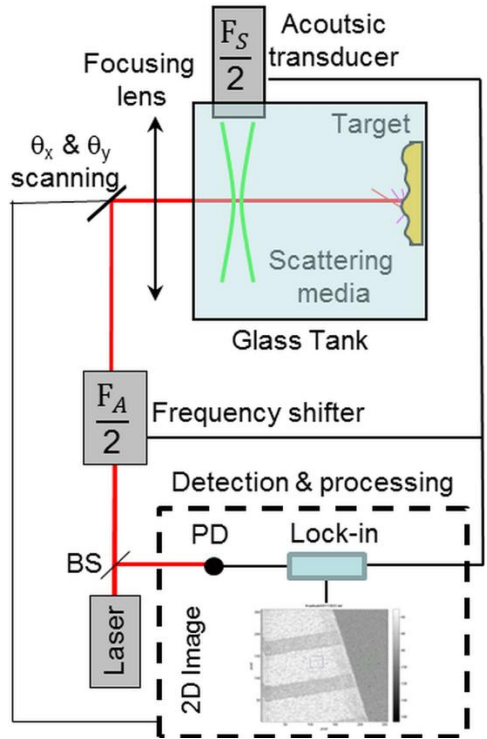

*Fig.1. S*chematic diagram of LOFI setup with acoustic tagging (AO-LOFI).PD, photodiode; BS, beam splitter.

In our experiment, the target is located inside a glass tank filled with a scattering medium. The input side of the tank produces a parasitic feedback which degrades significantly the SNR of the LOFI images. In these conditions, the "LOFI signal" corresponds to the coherent sum of the light back-scattered by the tank and by the target [8]. The noise equivalent power (NEP) is given by the sum of the quantum noise of the laser and the noise due to the tank parasitic feedback, we neglect the others parasitic backscattering. The SNR is then given by [7, 8]:

$$SNR = \frac{\sqrt{r_t^2 + r_P^2 + 2r_t r_P \cos\left(\frac{2\pi}{\lambda} 2(d_P - d_t)\right)} \cdot K_{opt} \cdot p_{out}}{\sqrt{\frac{2\Delta F}{R_{bs}}}\sqrt{p_{out}} + r_P K_{opt} \cdot p_{out}} \quad (1)$$

where $r_P$ and $r_t$ are the effective reflection coefficients of the input side of the tank and of the target respectively, $d_P$ and $d_t$ are the laser-tank and the laser-target distances respectively, $K_{opt}$ is the ratio between the light power at the input face of the tank and the power at the output of the laser, $\Delta F$ is the detection bandwidthand $R_{bs}$ is the beam splitter reflectivity.

When the incident power $P_{out}$ sent onto the target is increased in order to compensate for the scattering losses, the NEP is limited by the optical power back-reflected by the tank and the SNR shows an asymptotic behavior. The SNR is then intrinsically limited and there is no possibility to increase it.
In a LOFI setup with acoustic tagging (AO-LOFI), a US transduceris immerged in the tank (fig.1) to tag selectively by acousto-optic interaction, the photon that has traveled inside the tank. The selected transducer produces a focused US beam with a 1.75 mm diameter (at -6dB) at focus. The US frequency is noted $F_S/2$ and is equal to 2.25MHz . The transducer operating in the cw regime is placed and oriented in order to optimize the transverse overlapping between the laser beam and the focal zone of the acoustic wave. The transducer is located near the input side of the tank (close to the scanner) in order to keep a good overlapping during the scan of the target. Because of the short interaction length, the acousto-optical interaction is the Raman-Nath diffraction regime. The laser beam is split into three diffraction orders: 0, +1 and -1. The diffracted light is shifted in frequency by $+F_S$ and $-F_S$ for the order +1 and -1 respectively. The total frequency shift of the re-injected light is $F_A+F_S$, $F_A$ and $F_A-F_S$ for the diffraction orders +1, 0 and -1, respectively. The frequency shift of the parasitic feedback is always $F_A$consequently, it is possible to filter the parasitic feedback effect by selecting the frequency $F_A+F_S$ or $F_A-F_S$ as a reference for the lock-in detection. In these conditions, the AO-LOFI signal corresponds to the light back-reflected by the target, and the NEP is the quantum noise of the laser. The SNR is then given by [7]:

$$SNR = \sqrt{R_{bs}} \frac{r_t \kappa_{US} \cdot \gamma_{US}}{\sqrt{2\Delta F}} \sqrt{P_{out}} \quad (2)$$

where $K_{us}$ is the acousto-optic efficiency for the first diffraction order. Here, the SNR increases with the power sent on the target ($K_{opt}.p_{out}$) and with the integration time of the detection ($1/\Delta F$) and the NEP is shot noise limited.

### Experimental results
Two images obtained with the LOFI and the AO-LOFI setups are shown on figures 2a and 2b respectively, for comparison. The target is a metallic rule immersed inside a glass tank filled with water and a low milk concentration (2%). The corresponding scattering losses are about 5dB/cm (about 12 mean free paths on a round trip). The target is located 5cm away from the input side of the tank. The total round trip backscattering losses are then about 50dB. The RF power sent on the acoustic transducer is about 6.4W. The integration time and the frequency reference of the lock'in detection are 100µs and 3MHz respectively for both setups

On the images of fig.2 we can distinguish the edge of the ruler with two graduations. We can consider two distinct parts in these images: the left part corresponds to the ruler and allows to evaluate the LOFI or AO-LOFI signals, while the right part without the target allows to evaluate the so-called NEP. In each image, an average value of the SNR has been calculated for LOFI or AO-LOFI signal and NEP (see fig. 2). The ratio between the signal values in both images gives an acousto-optic efficiency of 13%. Despite a lower photometric budget, the SNR is higher in the AO-LOFI image than in the LOFI image. In our example, the corresponding SNR is about 5 with AO-LOFIand only 1.6 with the conventional LOFI. This difference can be explained by the isolation of the parasitic feedback produced by the tank in fig.2b.

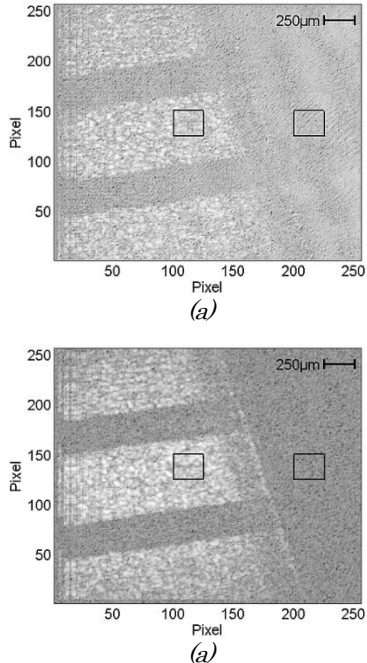

Fig.2. Images of the target in a scattering medium. a) with the LOFI setup at the $F_A$ reference frequency, $F_A$=3MHz. b) with the AO-LOFI setup at the $F_A$-$F_S$ frequency, $F_A$=7.5MH and $F_S$=4.5MHz. The left (resp. right) rectangle represents the area where the signal (resp. noise) is measured. Both rectangles represent 625 pixels.

In order to confirm this isolation effect, we have evaluated the SNR for both setups by varying the optical losses (i.e. the parameter $K_{opt}$ in eq. 1 and eq.2). The modulation has been obtained by a polarizer and a half-wave plate inserted between the beam splitter and the frequency shifter (fig.1). So as to realize images with a wide range of $K_{opt}$values, the scattering losses are lower than on fig.2, (about 2dB/cm), which explains a higher SNR. The shape of the curves presented on fig.3 is in good agreement with the predicted behavior by eq.1 and eq.2. The SNR has an asymptotic behavior versus K in LOFI images while it has a quasi linear evolution in AO-LOFI images. These results validate the parasitic feedback isolation in the AO-LOFI setup and the possibility to enhance the SNR with an increase of the integration time or with a higher laser output power.

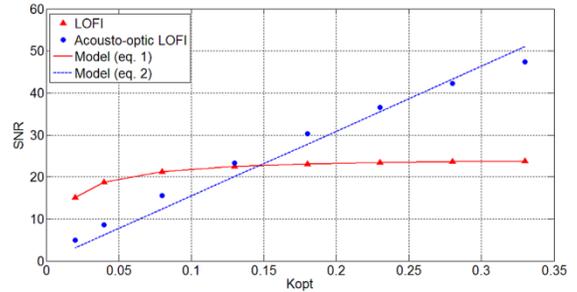

Fig.3. SNR versus parameter $K_{opt}$, in LOFI and AO-LOFI images.

## Conclusion

We have demonstrated the possibility to filter the parasitic feedback effect in LOFI image by acoustic tagging of the ballistic photons close to the studied target, which considerably enhances the sensitivity of the LOFI technique. Measurements validate the possibility to implement LOFI with a NEP shot noise limited: the detection limit is then given by the quantum noise of the laser. In a future work, we will use this tagging to enhance the sensitivity of the Synthetic Aperture LOFI technique [9] We will study also be the possibility to remove the frequency shifter made of two AOMs, which is needed in the setup depicted in fig. 1 to lift the degeneration of the frequency shift between the diffraction orders +1 and -1.